 \documentclass[journal, comsoc]{IEEEtran}

\usepackage{gensymb}
\usepackage{cite}
\usepackage{amsmath,amssymb,amsfonts}
\usepackage{graphicx}
\usepackage{xcolor}
\usepackage{kotex}  
\usepackage{caption}
\usepackage{subcaption}
\usepackage{caption}
\usepackage{subcaption}

\usepackage{stfloats}
\usepackage{float}
\usepackage{wrapfig}
\usepackage{xcolor}
\usepackage{colortbl} 
\usepackage{color,soul}
\usepackage{url}
\usepackage{verbatim}
\usepackage{graphicx}
\usepackage{cite}

\usepackage[acronym,toc,shortcuts]{glossaries}
\newacronym{BS}{BS}{Base Stations}
\newacronym{NTN}{NTN}{Non-Terrestrial Networks}
\newacronym{TN}{TN}{Terrestrial Networks}

\newacronym{NOMA}{NOMA}{Non-Orthogonal Multiple Access}
\newacronym{PD-NOMA}{PD-NOMA}{Power Domain Non-Orthogonal Multiple Access}
\newacronym{UEs}{UEs}{User Equipment's}
\newacronym{UE}{UE}{User Equipment}
\newacronym{UAV}{UAV}{Unmanned Aerial Vehicles}
\newacronym{SDG}{SDG}{Sustainable Development Goals}
\newacronym{UN}{UN}{United Nations}
\newacronym{HAP}{HAP}{High-Altitude Platforms}
\newacronym{AI}{AI}{Artificial Intelligence}
\newacronym{EE}{EE}{Energy Efficiency}
\newacronym{3GPP}{3GPP}{3$^{\text{rd}}$ Generation Partnership Project}
\newacronym{SA1}{SA1}{Service and System Aspects 1}
\newacronym{WG}{WG}{Working Group}
\newacronym{PLMN}{PLMN}{Public Land Mobile Networks}
\newacronym{MNO}{MNO}{Mobile Network Operators}
\newacronym{LEO}{LEO}{Low Earth Orbit}
\newacronym{KPI}{KPI}{Key Performance Indicators}
\newacronym{GEO}{GEO}{Geostationary Equatorial Orbit}
\newacronym{MEO}{MEO}{Medium-Earth Orbit}
\newacronym{CapEx}{CapEx}{Capital Expenditures}
\newacronym{IMS}{IMS}{IP Multimedia Subsystem}
\newacronym{SA}{SA}{System and Architecture}
\newacronym{RIS}{RIS}{Reconfigurable Intelligent Surfaces}
\newacronym{RAN}{RAN}{Radio Access Network}
\newacronym{NGSO}{NGSO}{Non-Geostationary Satellite Orbit}
\newacronym{UPF}{UPF}{User Plane Function}
\newacronym{ISL}{ISL}{Inter-Switch Link}
\newacronym{QoS}{QoS}{Quality-of-Service}
\newacronym{LAN}{LAN}{Local Area Network}
\newacronym{WLAN}{WLAN}{Wide Local Area Network}
\newacronym{ISAC}{ISAC}{Integrated Sensing and Communication}
\newacronym{SNO}{SNO}{Satellite Network Operators}
\newacronym{MENA}{MENA}{Middle East and North Africa}
\newacronym{LTE}{LTE}{Long-Term Evolution}
\newacronym{NAS}{NAS}{Non Access Stratum}
\newacronym{Wi-Fi}{Wi-Fi}{Wireless Fidelity}
\newacronym{FFR}{FFR}{Full Frequency Reuse}
\newacronym{ZF}{ZF}{Zero Forcing}
\newacronym{MIMO}{MIMO}{Multiple-Input-Multiple-Output}
\newacronym{MMSE}{MMSE}{Minimum Mean Square Error}
\newacronym{SSB}{SSB}{Synchronisation Signal Blocks}
\newacronym{CSI}{CSI}{Channel State Information}
\newacronym{CQI}{CQI}{Channel Quality Indicator}
\newacronym{RS}{RS}{Reference Signal}
\newacronym{GNSS}{GNSS}{Global Navigation Satellite System}
\newacronym{AMF}{AMF}{Access and Mobility Management Function}
\newacronym{LMF}{LMF}{Location Management Function}
\newacronym{5GC}{5GC}{5G Core Network}
\newacronym{C-JT}{C-JT}{Coherent Joint Transmission}
\newacronym{NC-JT}{NC-JT}{Non-Coherent JT}
\newacronym{MC}{MC}{Multi-Connectivity}
\newacronym{RRM}{RRM}{Radio Resource Management}
\newacronym{ML}{ML}{Machine Learning}
\newacronym{RU}{RU}{Radio Unit}
\newacronym{CU}{CU}{Control Unit}
\newacronym{DU}{DU}{Distributed Unit}
\newacronym{RIC}{RIC}{RAN Intelligent Controller}
\newacronym{CN}{CN}{Core Network}
\newacronym{EIRP}{EIRP}{Equivalent Isotropic Radiated Power}
\newacronym{VSAT}{VSAT}{Very Small Aperture Terminal}
\newacronym{FR}{FR}{Frequency Reuse}
\newacronym{CR}{CR}{Cognitive Radios}
\newacronym{DSA}{DSA}{Dynamic Spectrum Access}
\newacronym{gNB}{gNB}{Next-Generation Node B}
\usepackage{multirow}
\usepackage{array,boldline, makecell,booktabs}
\usepackage{longtable} 

\begin{document}
%
\title{Non-Terrestrial Networks for 6G: \\Integrated, Intelligent and Ubiquitous Connectivity}
%
%
%

\author{Muhammad Ali Jamshed,~\IEEEmembership{Senior Member,~IEEE,} Aryan Kaushik,~\IEEEmembership{Member,~IEEE,}  Miguel Dajer,~\IEEEmembership{Member,~IEEE,} \\Alessandro Guidotti,~\IEEEmembership{Member,~IEEE,} Fanny Parzysz,~\IEEEmembership{Member,~IEEE,} Eva Lagunas,~\IEEEmembership{Senior Member,~IEEE,} \\Marco Di Renzo,~\IEEEmembership{Fellow,~IEEE,} Symeon Chatzinotas,~\IEEEmembership{Fellow,~IEEE,} and Octavia A. Dobre~\IEEEmembership{Fellow,~IEEE} 
\thanks{
M. A. Jamshed is with the College of Science and Engineering, University	of Glasgow, UK (e-mail: muhammadali.jamshed@glasgow.ac.uk).\\  
$~~~$A. Kaushik is with the School of Engineering \& Informatics, University	of Sussex, UK (e-mail: aryan.kaushik@sussex.ac.uk).\\
$~~$M. Dajer is with Futurewei Technologies, USA (email: mdajer@futurewei.com).\\
$~~~$A. Guidotti is with the Consorzio Nazionale Interuniversitario per le
Telecomunicazioni, 43124 Parma, Italy (e-mail: a.guidotti@unibo.it).\\
$~~~$F. Parzysz is with the Orange Labs Network, France (email: fanny.parzysz@orange.com).\\
$~~~$E. Lagunas and S. Chatzinotas are with the Interdisciplinary Center for Security, Reliability and Trust (SnT), University of Luxembourg, Luxembourg (emails:\{eva.lagunas, symeon.chatzinotas\}@uni.lu).\\
$~~~$M. Di Renzo is with the Laboratory of Signals and Systems (L2S) of Paris-Saclay University, France. (email: marco.di-renzo@universite-paris-saclay.fr).\\
$~~~$O. A. Dobre is with the Faculty of Engineering and Applied Science, Memorial University, Canada (email: odobre@mun.ca).
}
}

%
%

\markboth{Submitted to IEEE Vehicular Technology Magazine}%
{Shell \MakeLowercase{\textit{et al.}}: Bare Demo of IEEEtran.cls for IEEE Journals}

\maketitle

\begin{abstract}
Universal connectivity has been part of past and current generations of wireless systems, but as we approach 6G, the subject of social responsibility is being built as a core component. Given the advent of Non-Terrestrial Networks (NTN), reaching these goals will be much closer to realization than ever before. Owing to the benefits of NTN, the integration NTN and Terrestrial Networks (TN) is still infancy, where the past, the current and the future releases in the 3$^{\text{rd}}$ Generation Partnership Project (3GPP) provide guidelines to adopt a successfully co-existence/integration of TN and NTN. Therefore, in this article, we have illustrated through 3GPP guidelines, on how NTN and TN can effectively be integrated. Moreover, the role of beamforming and Artificial Intelligence (AI) algorithms is highlighted to achieve this integration. Finally the usefulness of integrating NTN and TN is validated through experimental analysis.
\end{abstract}


\IEEEpeerreviewmaketitle


\section{Introduction} 
\label{sec:1}
Every generation of cellular communication is expected to surpass its predecessor by a fair margin, providing enhanced throughput, extremely low latency, seamless coverage, and ubiquitous connectivity. Among them, providing ubiquitous connectivity is still a challenging aspect, as there are still 2.6 billion people without mobile connectivity. Usually, the deployment of the cellular infrastructure directly relates to the population density, which in return limits the reach of advanced cellular technology to remote areas. Moreover, the cost of providing cellular services stays homogeneous throughout the whole country, which also limits the provisioning a comparable service to sparsely populated areas. The \ac{3GPP} Release 17 marks a significant milestone in overcoming these issues by integrating \ac{NTN} and \ac{TN}.

The term \ac{NTN} is often associated with satellite communications (always available \ac{NTN}) as the fight for space communications supremacy has captured everyone’s attention over the last few years. Other platforms that can provide communication services including \ac{UAV} and \ac{HAP} (on-demand \ac{NTN}), comes under the broader definition of \ac{NTN}. In general, \ac{NTN} provide connectivity in areas where \ac{TN} are unavailable or challenging to deploy, and are becoming mainstream as their role in supporting ubiquitous connectivity grows. There are a number of services that can be implemented utilizing \ac{NTN}, \emph{e.g.}, disaster recovery, search and rescue operations, and firefighting \cite{jamshed2024synergizing}. 

On one hand, the \ac{NTN} possess a significant importance in providing ubiquitous connectivity by integrating it with \ac{TN}. On the other hand, there are some challenges, such as delay due to extended path loss, Doppler shift, and interference management, which limits the seamless integration of \ac{NTN} and \ac{TN} \cite{toka2024ris}. Owing to these challenges associated with \ac{NTN}, this article presents an extensive overview of latest academic as well as industrial advancements in integrating \ac{NTN} and \ac{TN} to facilitate intelligent and ubiquitous connectivity. More specifically our contributions to knowledge are summarized as follows:

\begin{itemize}
    \item We provide a detailed overview of how the \ac{NTN} can be integrated with \ac{TN}, based on the \ac{3GPP} guidelines, covering the past, present and future releases. Moreover, we also highlight the key issues in the co-existence of \ac{TN} and \ac{NTN}, and introduce the \ac{3GPP}-based proposed solutions. 
    \item We provide a detailed overview of how user centric beamforming techniques can overcome the issues of inference in an integrated \ac{TN} and \ac{NTN}.
    \item We present a detailed overview of how the \ac{AI} algorithms can be exploited to perform successful integration of \ac{TN} and \ac{NTN}. Further we highlight key limitations of \ac{AI} for \ac{TN}/\ac{NTN} systems.
    \item We present two use case demonstrations of \ac{NTN}: 1) In an airborne scenario, \emph{i.e.}, by using \ac{UAV} to improve the \ac{EE} of \ac{UEs}. 2) In a satellite-based scenario, \emph{i.e.}, by using \ac{LEO} satellites to improve system capacity.
\end{itemize}

\section{NTN Ubiquitous Connectivity: A 6G Vision}

6G and \ac{NTN}-\ac{TN} integration are two emerging concepts in the realm of telecommunications and wireless technology, with some potential intersections and synergies. Although the final definition of 6G is still being formed, it is generally agreed that the social impact of 6G has to be a key tenet of the evolution of communications networks. A component of this social impact is universal broadband connectivity where the fusion of \ac{TN} and \ac{NTN} will play the enabling role. The integration of \ac{NTN} with 6G networks holds the promise of creating a highly connected world with ubiquitous, high-speed, and reliable communication services. In 5G the integration of \ac{NTN} has been at the higher layers; for 6G, the expectation is that this integration will be native.


\begin{figure*}
	\centering
	\includegraphics[width=0.9\textwidth]{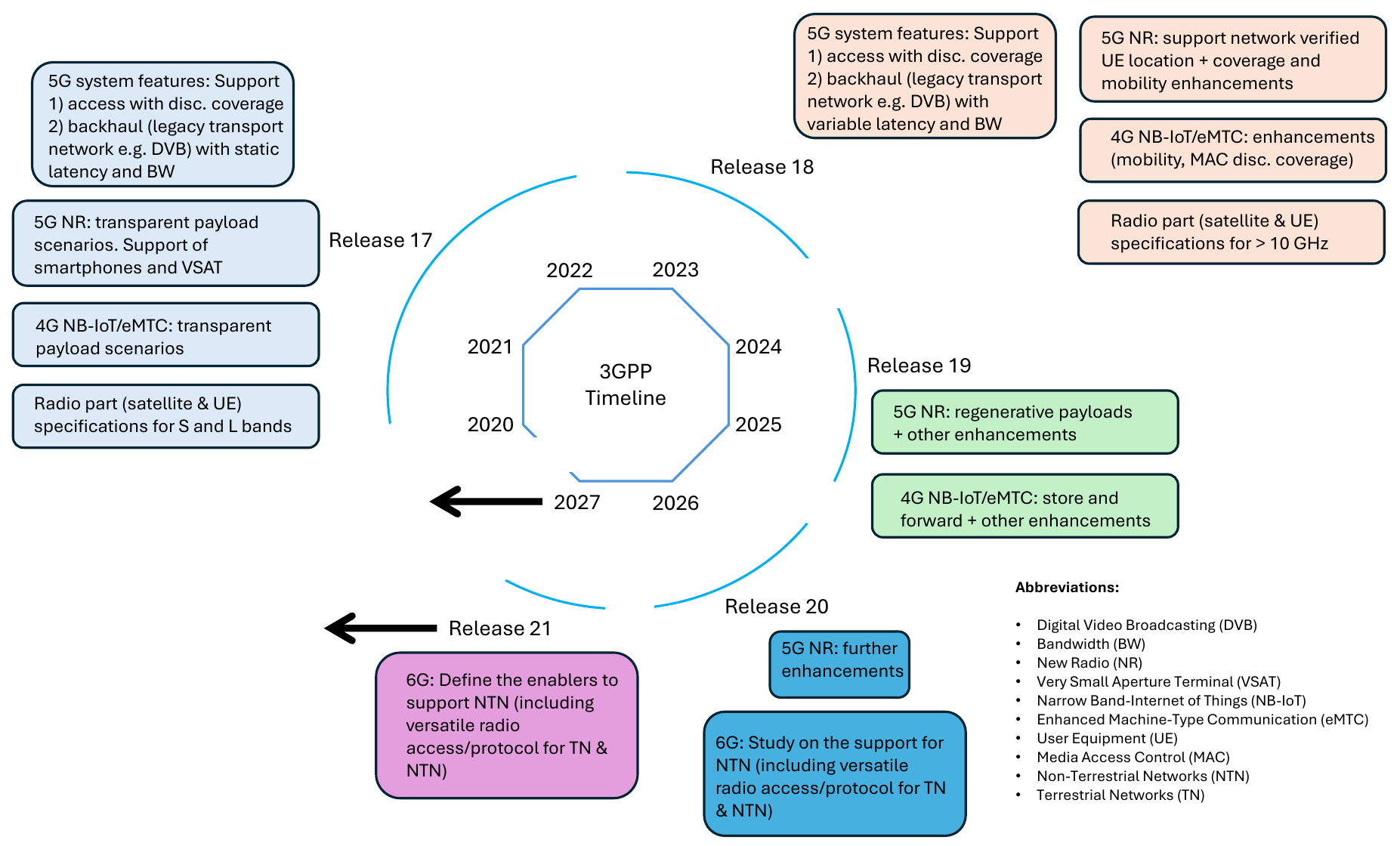}
	\caption{Timeline of the 3GPP standardization activities related to NTN.}
	\label{fig:3gpptimeline}
	\end{figure*}
 
The \ac{3GPP} standardization activities are represented in Fig.~\ref{fig:3gpptimeline}. Building on the success of Release 17, the work is now directed to the finalization of Release 18 and the beginning of Release 19, \emph{i.e.}, 5G-Advanced. In particular, regenerative payloads and operations in Ka-band are now being considered, in addition to different mobility and coverage enhancements. It is worthwhile mentioning that the activities in ITU-R for IMT-2023, not represented in the figure, have now started with the definition of the framework for \ac{TN}; the plan is to initiate the standard for terrestrial and non-terrestrial components in 2027. As for Release 20, \ac{3GPP} \ac{SA1} \ac{WG} has started the service requirement work. In this release, \ac{SA1} divides the work into 2 parts: part 1 will study and develop normative service requirement for the continued enhancement of 5G-Advanced; and part 2 will conduct a 6G study for preparing the normative work in Release 21, which will be the first \ac{3GPP} release for 6G. In \ac{3GPP} \ac{SA1} \#105 meeting held in May 2024 a new study item on satellite access \textemdash Phase 4 for Release 20 part 1 \textemdash was approved, the objective of this new study includes (S1-240312):

\begin{itemize}
    \item Enhanced support for emergency communications and mission critical services using satellite access.
    \item Support multi-orbits satellite access for multiple services.
    \item Support the ability to notify the user that a mobile terminated communication has failed when the \ac{UE} is unreachable in satellite access.
    \item Support \ac{IMS} voice calls using \ac{GEO} satellite access.
\end{itemize}

As for the part 2 of the 6G preparation, native support of both terrestrial and satellite network components to address a set of common goals and higher \ac{KPI} requirements for 6G (S1-233201) was proposed and getting more attention. It is still a bit early to arrive at a general conclusion on how \ac{NTN} will be integrated with \ac{PLMN} in \ac{3GPP}, but it is certain that these networks will be an integral part of the future communication infrastructure and of \ac{3GPP}’s 6G focus.

\subsection{NTN \& Standards}

Over the last couple of years, there have been several agreements between the traditional \ac{MNO} and \ac{SNO} aimed at providing extended service between the two networks. This arrangement allows \ac{MNO} to expand service coverage in places where the cost of deploying and operating a full terrestrial infrastructure may not be compensated by revenues. Satellite communications can be considered for home/office connectivity, backhauling of isolated \ac{BS}, or direct-to-smartphone connectivity. The first two scenarios are already quite widely deployed, \emph{e.g.}, in Latin America and \ac{MENA} countries, mostly based on proprietary solutions for satellite connectivity. The last scenario is also gaining attention, and the first compact terminals integrating the capabilities of smartphones and satphones have been launched on the market. \ac{3GPP} recognized the added value of the \ac{NTN}, and particularly satellite communication networks. This step forward in this ecosystem has raised significant interest. Indeed, it offers network interoperability, can favor the entry of new stakeholders in the market, limit vendor lock-in and reduce costs in a general manner. 

\ac{3GPP} targets the different scenarios for \ac{NTN}, from the support of non-\ac{3GPP} satellite access to fully \ac{3GPP}-compliant systems for 5G and beyond, and significant work has been pursued in this direction. Given this tie up between \ac{TN} and \ac{NTN}, it is critical that some level of standardization takes place. Starting with Release 17, \ac{3GPP} work in supporting \ac{NTN} has focused on ensuring seamless connectivity and interoperability between the two. Although \ac{NTN} includes other non-satellite-based (\emph{e.g.}, \ac{UAV} and \ac{HAP}) networks, \ac{3GPP} \ac{SA} \ac{WG}, so far, mainly focus on the satellite-based \ac{NTN}. 

\subsubsection{\textbf{Satellite Access Through Unmodified 4G Smartphones}}

In parallel with the \ac{3GPP} efforts towards standardized \ac{NTN}, several satellite companies, such as AST SpaceMobile, Lynk or Starlink, have demonstrated 4G capabilities, in partnerships with \ac{MNO}. Proprietary solutions have been developed to enable satellites to operate using the same \ac{3GPP} standards found in \ac{TN}, and transmit 4G \ac{LTE} signals from space to off-the-shelf smartphones, and vice versa. Initially, these services were limited to messaging type services, but performance is expanding to cover more advance voice and data capabilities. However, from a commercial perspective, such systems are not widely available yet. Firstly, this satellite connectivity needs to use spectrum in bands allocated to terrestrial mobile services and is thus subject to the approval of national regulatory authorities. In addition, further clarifications are still required concerning the sharing of responsibilities between the \ac{MNO} and the \ac{SNO}, to ensure the proper use of these frequency bands. Secondly, the interference issue remains a major one, to protect the terrestrial users of the same \ac{MNO}, as well as the cellular networks of neighboring countries. Indeed, the large footprint of satellite makes border management quite complex, especially in Europe, for example, where countries are generally smaller compared to North America. Finally, dedicating \ac{TN} bands to \ac{NTN} services is also a question of spectrum efficiency and business sustainability.

\subsubsection{\textbf{Integration of non-3GPP Satellite Access into 5G}}
\label{non3GPP}

The support of non-\ac{3GPP} access has already been specified in Release 15 to allow 5G to manage new types of access, with interoperability and reliability. To this end, different network functions have been defined, \emph{i.e.}, N3IWF (Non-3GPP Inter-Working Function) for untrusted access, and, TNGF (Trusted Non-3GPP Gateway Function), TWIF (Trusted \ac{WLAN} Inter-Working Function) and W-AGF (Wireline Access Gateway Function), for trusted access, depending on the considered \ac{UE} scenario and capabilities. These mechanisms have targeted primarily \ac{Wi-Fi}, fixed access, and mobile \ac{UEs} with limited support of 5G \ac{NAS} signaling, but adapting these functions to satellite-based scenarios has attracted significant interest \cite{MC1,MC2,MC3}. Moreover, there are numerous technical challenges, as highlighted in the projects \cite{INN3SCO, STARDUST_D52}. Indeed, it requires the reworking of these mechanisms to accommodate the specificities of space-to-ground links (\emph{e.g.}, long delays and lower capacity) and to adapt to various satellite protocol stacks, often based on proprietary technologies and generally tailored for specific use-cases.

\subsubsection{\textbf{5G-NTN: A First Complete Framework}}

\ac{3GPP} Release 16 is the first release in which \ac{3GPP} \ac{SA} \ac{WG}s tried to introduce the integration of satellite access technology into the overall 5G \ac{SA}. Unfortunately, in the Release 16 time-frame, only Stage 1 use cases and service requirement study (TR22.822) was completed, as well as the normative service requirements based on this study which are defined in TS22.261. The subsequent study for \ac{SA} level solution (TR 23.737) and the following normative standardization work based on the \ac{SA1} requirements were postponed and completed in Release 17. In Release 17, satellite \ac{NTN} was integrated into 5G \ac{TN} in the \ac{SA} with new 5G features covering the aspects of mobility management, incorporating the impact of delay, 5G system \ac{QoS} impact with satellite backhaul, and \ac{RAN} mobility with \ac{NGSO} regenerative-based satellite access.


\subsubsection{\textbf{NTN Systems in 5G-Advanced \& Beyond}} 
 
In Release 18, 5G \ac{SA} with satellite integration was continued to be studied and enhanced to cover satellite as an access technology (TR 23.700-28) and satellite as 5G system backhaul (TR23.700-27). The results of those two studies were incorporated into the final normative Release 18 Stage 2 specifications of TS23.501 and TS23.502. For the work on satellite-based \ac{NTN} as access technology, Release 18 work mainly focused on the mobility management enhancement and power saving for \ac{UE} with discontinuous satellite coverage. Moreover, for satellite-based \ac{NTN} as 5G system backhaul, new 5G solutions were developed to cover \ac{QoS} control enhancement considering dynamic satellite backhaul and using on board \ac{UPF} to provide support of satellite edge computing and local data switching to overcome the delay. Finally, the \ac{3GPP} Release 19 aims at studying key issues for satellite integration (TR23.700-29), such as providing support for regenerative-based satellites, store and forward satellite operations, and supporting communication between \ac{UEs} under the coverage of one or more serving satellites without the user plane traffic going through the ground network.

\section{{Advance Beamforming Techniques to Improve Future 6G NTN Connectivity }}

\begin{figure}
	\centering
	\includegraphics[width=\columnwidth]{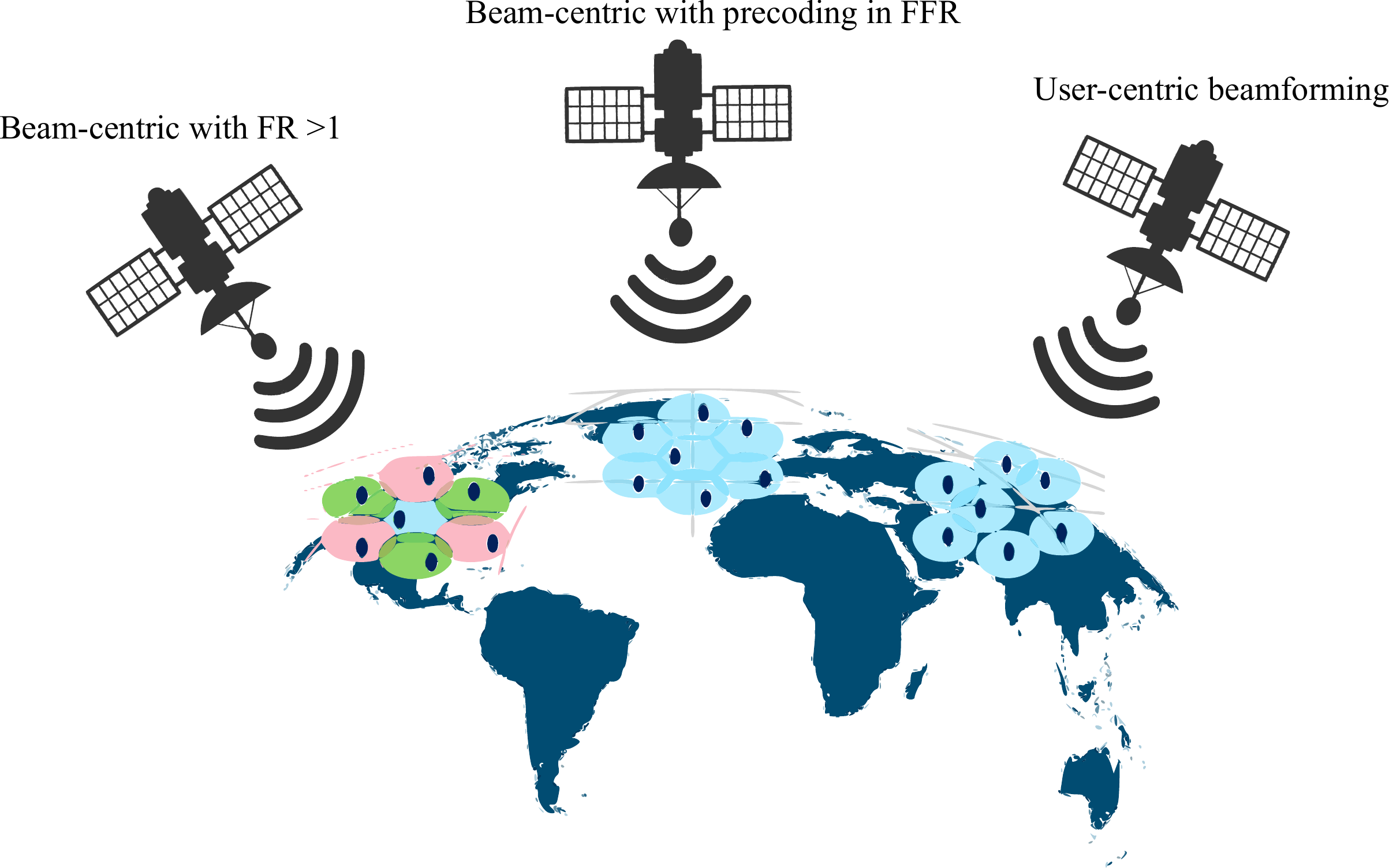}
	\caption{NTN beamforming and precoding schemes.}
	\label{fig:beamforming}
 \vspace{-0.5cm}
\end{figure}

Notably, legacy satellite communication systems are typically based on \ac{FR} schemes, for instance 3 or 4 colours, in which the available bandwidth is split into multiple non-overlapping channels differentiated in the frequency and polarization domains. Then, these channels are assigned to the on-ground coverage on a geographic basis to limit interference, by mapping them to specific beams of the desired beam lattice. Targeting higher capacities allowing to cope with the ever-increasing requirements for 5G, and now 6G. This can be achieved by means of advanced spectrum usage paradigms, such as \ac{CR}, \ac{DSA}, or by aggressively allocating the resources through \ac{FFR} schemes, in which the \ac{FR} factor is decreased to 1. Focusing on the latter approach, effective interference management techniques are required to cope with the massive co-channel interference, such as precoding. In this case, for a pre-determined on-ground beam lattice, \ac{MIMO} algorithms, such as \ac{ZF} or \ac{MMSE} are applied to a channel matrix at beam level. 

The implementation of these solutions via \ac{NTN} is already possible with the current Release 17-18 specifications, subject to a few conditions, such as: i) the number of simultaneous beams cannot exceed the number of \ac{SSB} that can be broadcasted by the \ac{gNB} (8 below 6 GHz and 64 above 6 GHz), allowing the \ac{UE} to identify and synchronize to the best serving beam \cite{38213}; and ii) the algorithms shall operate at beam level. In addition, one key aspect that still has to be addressed and agreed within \ac{3GPP} is related to beam management solutions, which define the mapping between the \ac{NTN} beams and the cell/NR beams managed by the \ac{gNB} \cite{38864}. Figure~\ref{fig:beamforming} depicts the different coverage solutions: beam-based with \ac{FR} schemes, beam-based with precoding supporting full \ac{FR}, and user-centric beamforming, in which the beams are generated in the direction of the users instead of being determined a priori on a geographic basis.

Beam-level solutions, such as precoding, show a good performance, but do not yet exploit the full potential of \ac{MIMO} via \ac{NTN}. In fact, the beam lattice is usually defined to geographically limit co-channel interference, rather than exploiting it. In this framework, user-centric beamforming, \emph{i.e.}, beamforming algorithms that dynamically generate beams in the direction of the served users, are recognised as the next evolution; however, these are not yet defined at standard level. 

A first aspect to be taken into account is related to the ancillary information that is exploited to compute the beamforming coefficients, which classifies the user-centric algorithms as \cite{beamforming_1}: i) \ac{CSI}-based, when the channel coefficients are estimated by the users and reported to the network; and ii) location-based, in which the users estimate their location and then, the network infers the channel coefficients exploiting this information and the \ac{NTN} node ephemeris. The implementation of \ac{CSI}-based solutions requires that each user is able to estimate the channel coefficient between its receiving antenna and each radiating element on-board. Currently, \ac{CSI} estimation in \ac{3GPP} \ac{NTN} allows to estimate a quantized information on the \ac{CQI} at beam level, by exploiting the \ac{CSI}-\ac{RS} defined in the \ac{3GPP} specifications. Thus, to support these techniques, future releases shall provide the means to implement pilot-aided estimation algorithms at radiating element level, which might increase the signalling overhead. This is a challenging task, in terms of both computational complexity (the number of radiating elements is usually large, \emph{e.g.} 512 or 1024) and of signalling overhead. A possible solution is that of exploiting the current \ac{CSI}-\ac{RS} to estimate the common portion of the channel coefficients (everything happening between the \ac{NTN} node and the terminal), while leaving at the \ac{gNB} the task of superimposing the terms related to the known antenna array geometry. As for location-based solutions, in 5G the users are equipped with \ac{GNSS} capabilities and they can report their location; however, the location information is not available at \ac{RAN} level, \emph{e.g.}, at the on-board \ac{gNB}, but in the \ac{5GC}. More specifically, the \ac{AMF} interacts with the \ac{LMF} to provide location services. Thus, also location-based solutions would need adaptations, in order to make the location information available in the \ac{RAN} and not only in the \ac{5GC}. 

More recently, the implementation of Cell-Free \ac{MIMO} via \ac{NTN} is also receiving an increasing attention, \cite{beamforming_1}. In this scenario, multiple close \ac{NTN} nodes organized in a flying swarm can cooperate to generate a virtual antenna array in space, aimed at implementing more flexible and advanced user-centric beamforming solutions. In addition to the above-mentioned evolution's required to support also standalone beamforming, the major challenge is related to the tight time and frequency synchronization requirements on the \ac{ISL}. In fact, for distributed (or federated) user-centric beamforming solutions to be effective, and not detrimental to the performance, the non co-located transmissions from the \ac{NTN} segment shall be ideally perfectly aligned in time and frequency. Notably, this requirement is particularly challenging to be achieved. This is one of the main reasons for which these solutions, denoted as \ac{C-JT} were de-prioritized in favour of \ac{NC-JT} approaches such as multiple transmission/reception point or \ac{MC} in past approaches.


\section{AI for Seamless Connectivity in 6G NTN}


The seamless integration of \ac{NTN} into 6G networks provides another level of complexity to an infrastructure that is already characterized by dense \ac{BS} deployment, with an exponentially increasing number of terminals, each with heterogeneous \ac{QoS} requirements and random mobility patterns. The conventional model-based design cannot deal with such complex network management, as they usually lack mathematical tractability and involve significant computational cost. In this context, data-based tools can provide speed-up inference procedures by learning the patterns and relationships of complex algorithms, assuming they have been previously trained with relevant datasets. Offline training is often complemented by updating the \ac{AI} model based on new evidence, which can significantly improve the adaptability of the model to changes on the environment of operation.

The \ac{3GPP} Release 18, which is the first standard of 5G-Advanced, includes the discussion of a suitable framework for the integration of \ac{AI} in the NR air interface targeting automation of networks \cite{XingqinLin23}. Few study items identified areas and use-cases where \ac{AI} can definitely have an impact. These are: (i) \ac{AI} for network management and orchestration; (ii) \ac{AI}-enabled \ac{RAN} intelligence (with a focus on energy efficiency); and (iii) \ac{AI}-native air interface (with different collaboration levels between \ac{gNB} and \ac{UE}), including \ac{CSI}, and beam management and positioning. \ac{3GPP} has also identified procedures to perform the data analytics function and interact with the \ac{5GC} (\emph{i.e.}, 5G network data analytics function).

In the next sub-sections, we first present the applications of \ac{AI} in integrated 6G-\ac{NTN} networks and subsequently, we discuss the related challenges and limitations.

\subsection{AI Applications in Integrated 6G-NTN}

6G-\ac{NTN} optimization and management is one of the areas with highest potential for \ac{AI}-oriented solutions. With a combination of ground \ac{BS} and multiple flying satellites with overlapping coverage, the space and ground operations need to be strictly coordinated for seamless service towards their users. The network shall be optimized such that the resource efficiency is maximized and the service agreements with customers are met. \ac{AI} provides the solution to flexible and autonomous network adaptation to wireless channel changes, rapid traffic demand variations, as well as mobility patterns. The most common approach is to use deep learning architectures to emulate computational-stringent algorithms, in an attempt to find a suitable trade-off between performance and speed of convergence. If environment testing is allowed, reinforcement learning can be considered, where different configurations are tested in a trial-and-error fashion until convergence. 

\begin{figure}[!t]
\centering
\includegraphics[width=\columnwidth]{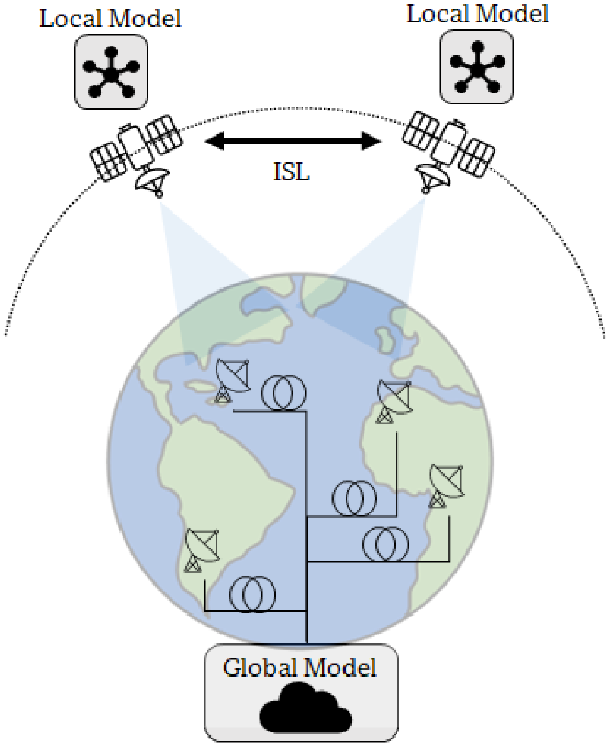}
\caption{Schematic of a distributed AI model over a constellation of satellites.}
\label{AI_Figure_fig}
\end{figure}

As discussed, pro-activity to network congestion and failures is critical, and requires continuous monitoring and analysis of network performance data. \ac{AI}-tool devoted to anomaly detection (\emph{e.g.}, interference and/or link failure detection) as well as prediction of network metrics (\emph{e.g.}, demand, trajectory of terminals, congestion) have been shown to provide an advantage to network management procedures, allowing to take corrective actions before the users' \ac{QoS} starts dropping. Such procedures become relevant in \ac{NTN} scenarios, where the spectrum is congested due to the many \ac{LEO} satellites being launched, and interference events are likely to occur. Link congestion is also a characteristic of \ac{NTN}, as the wide coverage area of a single satellite encompasses many more users than a terrestrial \ac{BS}. Furthermore, the fact that satellites revisit the same area several times can be exploited as depicted in Fig. \ref{AI_Figure_fig}, where the satellite \ac{AI} local models are collected at a cloud via optical interconnected ground stations and updated based on shared experience. \ac{ISL} may become useful in forming cluster-based models with nearby satellites, \emph{i.e.}, by exploiting the experience of neighboring satellites to build stronger models.


\subsection{Limits of AI-enabled TN/NTN Systems}

One of the key aspects of the O-RAN architecture is that it is natively designed for \ac{AI}-empowered \ac{RAN}. \ac{RRM} can be optimized at the near-real-time scale ($<$ 1sec) for efficient \ac{gNB} radio configuration, and at the non-real-time scale (minutes or hours) for dynamic \ac{TN}/\ac{NTN} load balancing (including management of satellite beam \ac{EIRP} or on-board data routing). To this end, the O-RAN \ac{ML} framework consists in: 1) several interfaces to collect data at the \ac{gNB} (\ac{RU}/\ac{CU}/\ac{DU}/\ac{RIC}), the \ac{UE}, the \ac{CN}, but also different application functions; 2) the \ac{ML} training host for online and offline training; and 3) the \ac{ML} interference host, used in particular for execution, including hosting the \ac{ML} model and providing outputs to the actor of the \ac{ML}-assisted solution.

Firstly, the location of these \ac{ML} model components has significant impact on the system performance but \ac{NTN} raises specific challenges, yet to be fully explored. Different functional splits can be envisaged for \ac{NTN}, \emph{e.g.}, the transparent mode (full \ac{gNB} on the ground), but also \ac{RU} on board, \ac{RU}/\ac{DU} on board or full \ac{gNB} on board. Each case implies different practical constraints for deploying and running the \ac{ML} framework. The right balance still needs to be found between long propagation delays (compared to the \ac{TN} scenario), potentially high number of served \ac{UEs} (due to the large satellite footprint) and above all, the limited computational capabilities of the satellite payload. Furthermore, the moving topology of \ac{LEO} mega-constellations is another major challenge to practical implementation. Secondly, the \ac{AI} efficiency may be restricted by the availability, quantity and quality of data. Data is often collected at different entities (\ac{UEs} or network) in a fully distributed manner and at very different space/time scales, such that the current O-RAN architecture, interfaces, and orchestration capabilities require adaptation. By introducing new \ac{AI} generative models, the lack of data availability may be overcome by generating synthetic entries that emulate realistic environment. Traffic modeling and prediction is another key aspect shaping \ac{ML} models for optimized resource management and will significantly impact the choice of the best option for \ac{AI}-empowered \ac{TN}/\ac{NTN} system. 

Finally, a large number of \ac{AI} algorithms proposed in the literature should be principally considered as theoretical (yet valuable) upper bounds, due to the underlying assumption that the whole \ac{TN}/\ac{NTN} system is operated by a single stakeholder. Due to their different technical architecture, historical ecosystem evolution, regulation landscape and geographical footprint, \ac{TN} and \ac{NTN} will probably be organized differently from a business perspective. Main solutions for \ac{TN}/\ac{NTN} interconnections are roaming, multi-connectivity with converged core (as detailed in Section \ref{non3GPP}), \ac{RAN} sharing and all their variants. Each case corresponds to a different level of network interworking, with limited information sharing, such that part of data sources may remain inaccessible, due to confidentiality. Investigating the gap between theoretical upper bounds (with full data knowledge) and constrained \ac{ML} frameworks could provide a good performance indicator of \ac{AI} applied to practical \ac{TN}/\ac{NTN} architecture options and business models.

\begin{figure*}[!t]
     \centering
     \begin{subfigure}[t]{0.5\textwidth}
         \centering
         \includegraphics[width=\columnwidth]{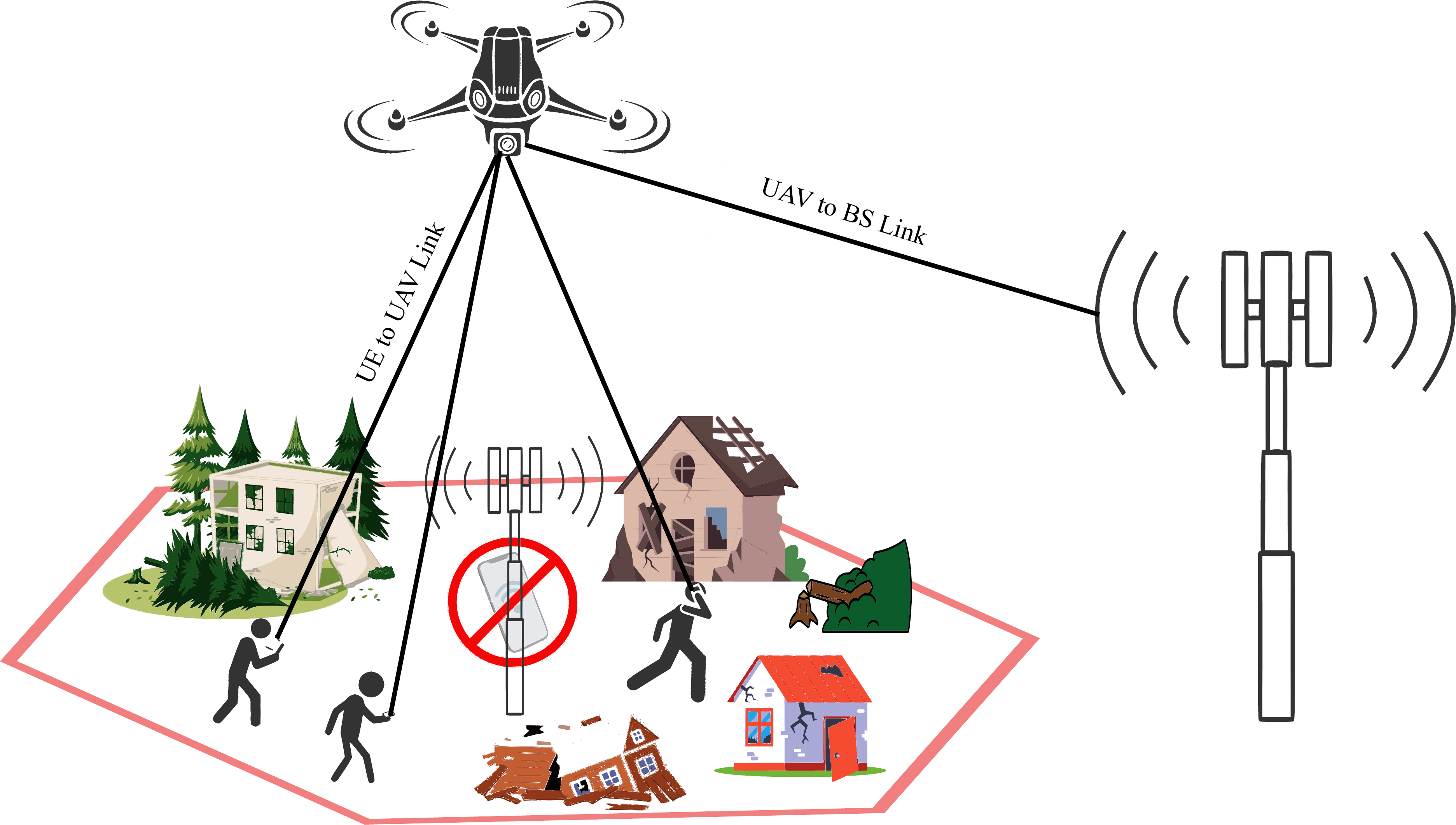}
         \caption{System model, where a UAV is employed for disaster management.}
         \label{fig1}
     \end{subfigure}
     \hfill
     \begin{subfigure}[t]{0.45\textwidth}
         \centering
         \includegraphics[width=\columnwidth]{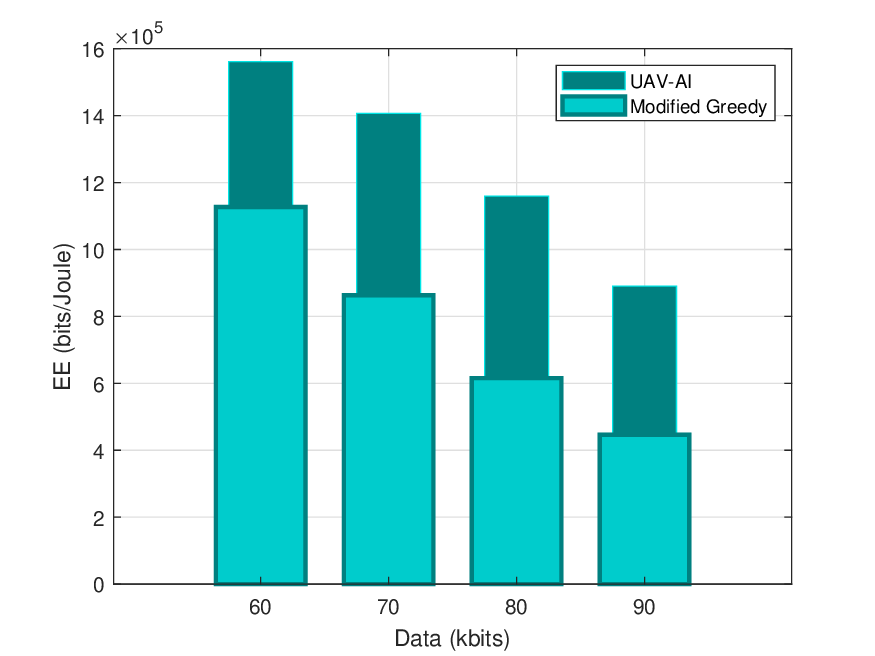}
         \caption{EE versus the transmitted data size for a fixed number of UEs.}
         \label{fig2}
     \end{subfigure}
        \caption{Public safety use case demonstration of NTN, jointly using AI and NOMA.}
\end{figure*}

\section{NTN Design and Analysis}
\subsection{Airborne NTN Connectivity}
In this section, we have demonstrated through simulations, a public safety use case of \ac{NTN}. As illustrated in Fig. \ref{fig1}, we have considered a scenario where the \ac{UEs} are unable to communicate with the \ac{BS} due to a natural disaster. We have optimized the limited energy levels of \ac{UEs} available in the disaster region, by relying on \ac{NOMA}, \ac{AI} and a \ac{UAV}. As presented in \cite{jamshed2021unsupervised}, firstly the usefulness of \ac{NOMA} is exploited by performing user grouping and subcarrier allocation of devices by collectively using the k-means, F-test and elbow methods. Afterwards, power allocation is performed using iterative methods, to improve the \ac{EE} of \ac{UEs}. We have considered 128 subcarriers, a bandwidth of 10 MHz and the path loss model defined in \cite{wu2016green}. In Fig.~\ref{fig2}, \ac{EE} achieved by the UAV-AI solution is illustrated with varying the data transmitted, whereas the uplink power level, circuit power and number of \ac{UEs} are fixed to 0.2 Watts, 1.4002, and 70. In general, by increasing the amount of data transmitted for a fixed number of \ac{UEs}, the total \ac{EE} decreases, which shows a proper functioning of the UAV-AI solution. In comparison to a modified version of the greedy algorithm, which does not maintain fairness among all the \ac{UEs}, the UAV-AI algorithm improves the \ac{EE} consistently for all amounts of data transmitted. Approximately, the performance gap remains the same throughout in Fig.~\ref{fig2}, which is due the fixed circuit power value. 


\begin{figure*}
     \centering
     \begin{subfigure}[t]{0.5\textwidth}
         \centering
	\includegraphics[width=\columnwidth]{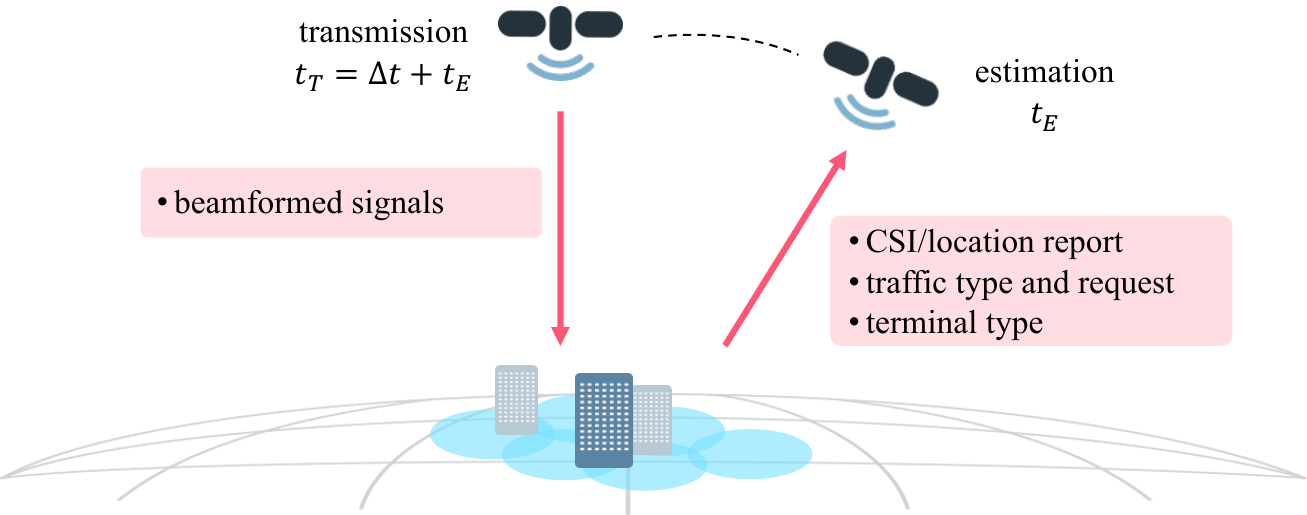}
	\caption{System model for user-centric beamforming.}
	\label{fig:beamforming_2}
     \end{subfigure}
     \hfill
     \begin{subfigure}[t]{0.45\textwidth}
         \centering
	\includegraphics[width=\columnwidth]{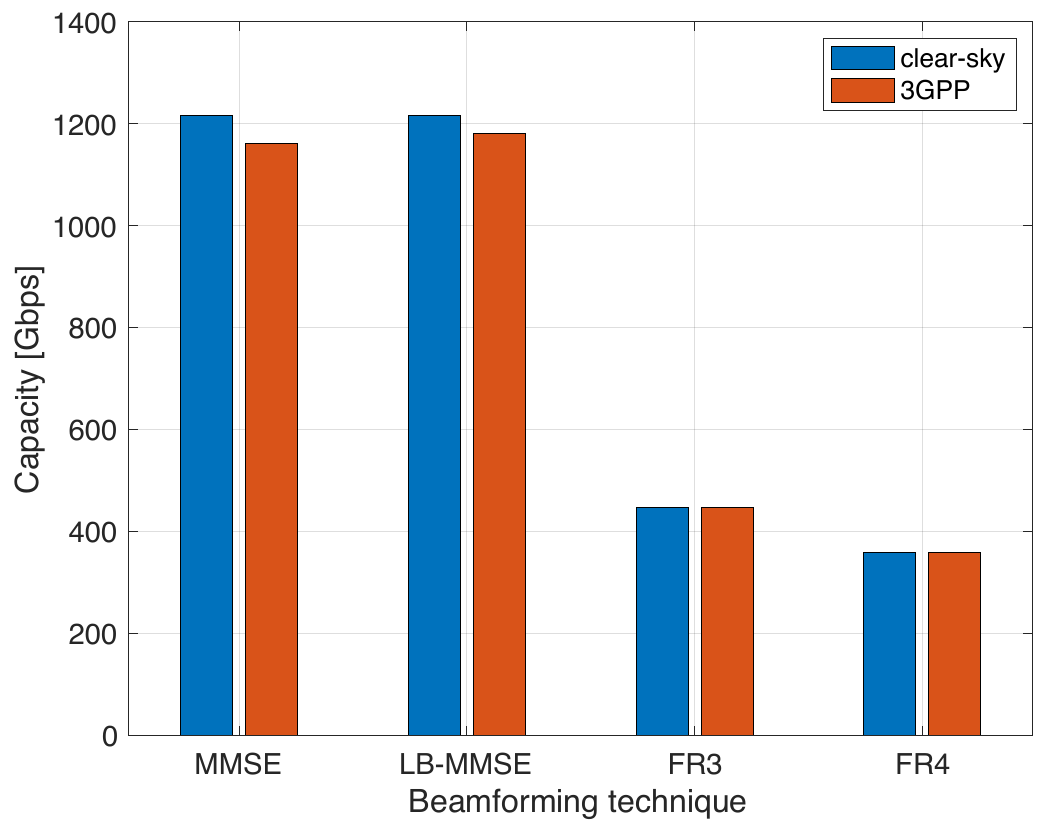}
	\caption{System capacity with user-centric beamforming.}
	\label{fig:bf_capacity}
     \end{subfigure}
             \caption{Illustration of a user centric beamforming use case, while utilizing LEO satellites to improve the system capacity.}
\end{figure*}



\subsection{LEO Satellites based NTN Connectivity}
In this section, we show the numerical results for the implementation of user-centric beamforming via \ac{NTN} from a single \ac{LEO} satellite. The antenna model and system configuration are detailed in \cite{beamforming_1}. Fig.~\ref{fig:bf_capacity} shows the system capacity achieved with an available transmission power on-board of $38$ dBW, transmitting in the Ka-band at $20$ GHz and assuming a $400$ MHz bandwidth (the maximum allowed for \ac{NTN} in Ka-band) for \ac{VSAT} receivers. The beamfomring algorithms are \ac{CSI}-based \ac{MMSE} and Location-Based MMSE (LB-MMSE) and their performance is compared to legacy beam-based schemes with FR3 and FR4. For MMSE and LB-MMSE, maximum power constraint is considered as the power distribution strategy, which is the one guaranteeing that the orthogonality in the beamforming vectors is maintained and the on-board power limitations are met. Two channel models are considered: clear-sky and \ac{3GPP}, which identifies the system-level channel model detailed in \ac{3GPP} TR 38.811, \cite{38811} and TR 38.821, \cite{38821}. As shown in Fig.~\ref{fig:beamforming_2}, during the estimation phase, the users estimate their \ac{CSI} or location to allow the \ac{gNB} to compute the beamforming matrix; however, the actual transmission occurs after a misalignment time $\Delta t$: the longer this interval, the less accurate the ancillary information is for user-centric beamforming. In the considered example, $\Delta t=16.7$ ms, computed assuming a \ac{LEO} satellite at $600$ km and minimum elevation angles to the users and the gateway of $30^{\circ}$ and $10^{\circ}$, respectively.

It can be noticed that the advantage in applying user-centric solutions is massive, with a relative gain of approximately $160-170$\% and $230-240$\% compared to FR3 and FR4, respectively. In general, the achievable spectral efficiency is larger with FR3 and FR4, thanks to the reduced co-channel interference; however, the exploitation of the entire bandwidth through full \ac{FR} with beamforming yields such large gains. It can also be noticed that, in the \ac{3GPP} channel that considers additional losses (scintillation, shadowing, atmospheric loss), LB-MMSE performs better than the \ac{CSI}-based solutions. This is in line with the considerations reported in \cite{beamforming_1}. In fact, location-based techniques infer the channel coefficients based on the users' positions and do not take into account additional losses; \ac{CSI}-based algorithms take the entire estimated channel, which is however different from the channel encountered when transmitting the signals and this misalignment degrades the performance.

\label{sec:5}

\section{Conclusion}
The integration of \ac{NTN} and \ac{TN} will play a key role in fulfilling the ITU 2030 objective of providing universal access and connectivity. This integration still faces numerous open research questions and challenges that are yet to be addressed. In this article, we have provided an overview of how \ac{3GPP} is making efforts towards addressing such issues. We have also showcased how \ac{AI} and beamforming techniques can play a role in achieving this successful integration. Finally, through experimental validation, first we have demonstrated on how airborne \ac{NTN} connectivity can improve the \ac{EE} of \ac{UEs} in a disaster region, and then, how the use of user-centric beamforming via satellites can improve the total system capacity of \ac{TN}.


\bibliographystyle{IEEEtran}

\bibliography{IEEEabrv,BibRef}


\end{document}